\documentclass[aps,twocolumn,superscriptaddress]{revtex4-2}

\usepackage{amsfonts}
\usepackage{amsmath}
\usepackage{amssymb}
\usepackage{graphicx}
\usepackage{bm}
\usepackage{xcolor}
\usepackage{array}

\renewcommand{\d}{\text{d}}

\begin{document}

\title{Competition between thermocapillary and solutocapillary flows  in thin liquid films}

\author{Darsh Kumar}
\affiliation{Department of Mechanical Engineering,
IIT Kanpur, Kanpur, Uttar Pradesh 208016, India}

\author{Pradipta Kumar Panigrahi}
\affiliation{Department of Mechanical Engineering,
IIT Kanpur, Kanpur, Uttar Pradesh 208016, India}

\author{Thomas Bickel}
\email{thomas.bickel@u-bordeaux.fr}
\affiliation{Univ. Bordeaux, CNRS, Laboratoire Ondes et Mati\`ere d'Aquitaine, 33400 Talence, France}

\date{\today}
\begin{abstract}
We investigate the thermocapillary flow in a thin liquid film which is subjected to local heating, in the presence of insoluble surfactants. 
While surfactant molecules are first advected from warmer  to cooler regions, the resulting concentration gradient drives a solutal counterflow in the opposite direction. This competition is theoretically addressed within the lubrication approximation. Assuming small deviations with respect to the mean surfactant concentration, we derive the time evolution equation governing the shape of the interface. Our study reveals that both interfacial deformations and velocities are progressively suppressed as the solutal Marangoni number increases. Our versatile model, adaptable to a range of experimental setups, offers a quantitative tool for understanding the effect of surfactants in thermocapillary-driven systems.
\end{abstract}

\maketitle

\section{Introduction}

The ability to control the topography of a thin liquid film is of paramount importance for a wide range of technological processes~\cite{brunPRFluids2024}. One effective method for inducing surface deformation is by generating a flow in the liquid phase through capillary forces~\cite{oronRMP1997}. Indeed, the surface tension of the liquid-air interface can be manipulated using for instance electric or magnetic fields, concentration gradients, as well as thermal gradients. In particular, the thermocapillary effect has gained significant attention in recent years due to its  remarkable versatility~\cite{darhuberARFM2005,karbalaeiMuMa2016,chenChemRev2022,vermaLSA2022}. For most simple liquids, the surface tension is a decreasing function of the temperature. A tangential temperature gradient then induces a shear stress along the interface, and therefore a flow in the liquid phase. Thermocapillary convection has for instance been exploited in order to shape liquid films with a predetermined topography~\cite{rubinLSA2019,eshelFlow2022,ivanovaColInt2022} or to tune the thickness of ultra-thin films  down to a few nanometers~\cite{clavaudPRL2021}. 
The thermocapillary effect has also been proven effective for contactless actuation of microparticles~\cite{maggiNatComm2015,girotLangmuir2016,haywardPNAS2021} and for the manipulation of  flows in liquid droplets~\cite{baroudLabC2007,yakshiAPL2010,pradhanExpFluids2015}. 

Although thermocapillary convection is well established, it is widely acknowledged that the purity of the system is critical~\cite{levichbook}. This point is problematic when working with high-surface-energy liquids such as water, that are prone to contamination by the surface-active molecules that are invariably present in the surrounding environment~\cite{bezuglyiCSA2004}. Even at trace concentrations, surfactant molecules can profoundly alter the hydrodynamic behavior of the liquid-air interface.
Indeed, the molecules are transported from the warmer regions to the cooler regions by the thermocapillary flow, resulting in their depletion in the former and accumulation in the latter. This generates a concentration gradient that drives a solutal counterflow toward the regions of lower concentration, \textit{i.e.}, from colder  to warmer regions. This secondary flow can then partially or fully counteract the original thermocapillary flow~\cite{homsyJFM1984,carpenterJFM1985,shmyrovJFM2019}. Consequently, the hydrodynamic boundary condition may shift from a no-stress to a no-slip condition due to the presence of even a small amount of contaminants.

Several recent studies have sought to elucidate the intricate interactions between thermal, viscous, and molecular transport processes~\cite{pinanPoF2021,rudenkoJFM2022,vinnichenkoIJHMT2018,bickelEPJE2019,koleskiPoF2020}, but a comprehensive understanding of these coupled phenomena remains elusive. In an effort to address this gap, a systematic approach was recently developed to characterize the behavior of these otherwise undetectable impurities~\cite{bickelPRFluids2024}. This work emphasized the critical influence of the solutal Marangoni number in governing the transport dynamics. However, the model was limited to the scenario of a deep liquid layer, overlooking the deformations of the interface. The objective of the present study is to extend this analysis to thin-film geometries, where interfacial deformations are expected to become significant.

The remaining of the paper is organized as follows. The governing transport equations are presented in Section~\ref{sec_lubrication}. In Section~\ref{sec_thermocap}, we explicitly derive the surface tension gradient for a pure thermocapillary flow, while Section~\ref{sec_solutocap} extends the analysis to the case where both thermocapillary and solutocapillary effects are present. Section~\ref{sec_deform} examines the impact of the competition between thermocapillary and solutocapillary flows on the interface shape. Finally, a summary of the results and a discussion of their implications is provided in Section~\ref{sec_disc}.

\section{Marangoni flow in the lubrication approximation}
\label{sec_lubrication}

\begin{figure}
\centering
\includegraphics[width=\columnwidth]{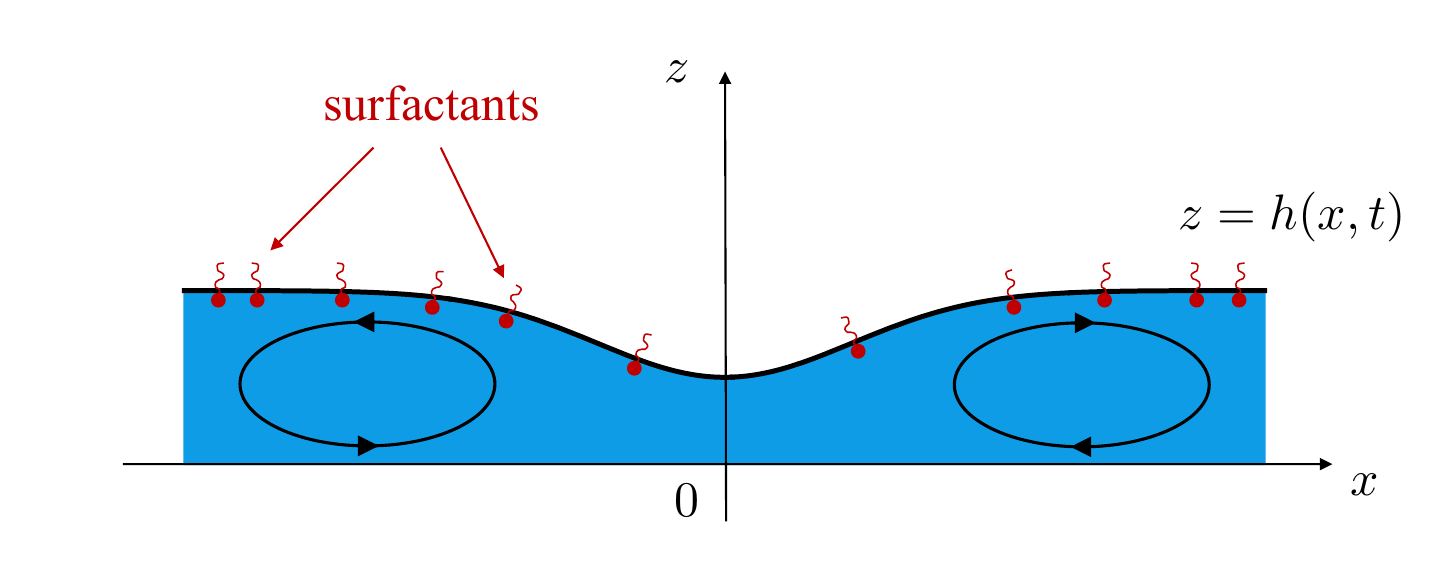}
\caption{Schematic of the problem. A thin liquid film is deformed due to the original thermocapillary flow. If the interface is loaded with surfactants, the flow and therefore the overall shape of the interface are profoundly modified.}
\label{fig1}
\end{figure}

\begin{table}
\begin{ruledtabular}
\begin{tabular}{l  c c}
Symbol & Definition & SI unit\\
\hline
$x,y,z$ & Cartesian coordinates & $\text{m}$ \\
$v_x,v_y,v_z$ & Velocity components & $\text{m}\,\text{s}^{-1}$ \\ 
$v_s$ & Interfacial velocity & $\text{m}\,\text{s}^{-1}$ \\
$p$ & Pressure & $\text{Pa}$ \\
$\bm{\sigma}$ & Stress tensor & $\text{Pa}$ \\
$\eta$ & Dynamic viscosity & $\text{Pa}\,\text{s}$ \\
$\rho$ & Mass density & $\text{kg}\,\text{m}^{-3}$ \\
$\mathbf{g}$ & Gravitational acceleration & $\text{m}\,\text{s}^{-2}$ \\
$\Gamma$ & Local concentration of surfactants & $\text{m}^{-2}$ \\
$\Gamma_0$ & Equilibrium concentration & $\text{m}^{-2}$ \\
$D_s$ & Diffusion coefficient of surfactants & $\text{m}^2\,\text{s}^{-1}$ \\
$T$  & Local temperature & $\text{K}$ \\
$T_0$ & Equilibrium temperature & $\text{K}$ \\
$\Delta T$ & Temperature variation & $\text{K}$ \\
$T_s$ & Interfacial temperature & $\text{K}$ \\
$c$ & Heat capacity & $\text{J}\, \text{kg}^{-1}\,\text{K}^{-1}$ \\
$\kappa$ & Heat conductivity & $\text{W}\, \text{m}^{-1}\,\text{K}^{-1}$  \\
$\alpha = \kappa/(\rho c)$ & Heat diffusion coefficient & $\text{m}^2\,\text{s}^{-1}$ \\
$k$ & Heat transfer coefficient & $\text{W}\, \text{m}^{-2}\,\text{K}^{-1}$ \\
$\gamma$ & Surface tension & $\text{J}\, \text{m}^{-2}$  \\
$\gamma_0=\gamma(T_0)$ & Equilibrium surface tension & $\text{J}\, \text{m}^{-2}$  \\
$\gamma_{\theta}=\vert \partial \gamma/\partial T \vert$ & Thermocapillary coefficient & $\text{J}\, \text{m}^{-2} \, \text{K}^{-1}$  \\
$\gamma_{s}= \vert \partial \gamma/\partial \Gamma \vert$ & Solutocapillary coefficient & $\text{J}$  \\
$h$ & Local film thickness & m \\
$h_0$ & Equilibrium film thickness & m \\
$\ell_{c}=\sqrt{\gamma_0/(\rho g)}$ & Capillary length & m \\
$\ell_{\theta}$ & Thermal  length  & m \\
$X=x/\ell_c$ & Dimensionless horizontal position & -- \\
$Z=z/h_0$ & Dimensionless vertical position & -- \\
$H=h/h_0$ & Dimensionless film thickness & -- \\
$\lambda_c =\ell_c/\ell_{\theta}$ & Dimensionless capillary length & -- \\
$\varepsilon=h_0/\ell_{\theta}$ & Lubrication parameter & -- \\
$U_{\theta}=\gamma_{\theta} \Delta T / \eta$ & Thermocapillary velocity  & $\text{m}\,\text{s}^{-1}$ \\
$U=\varepsilon U_{\theta}$ & Characteristic velocity & $\text{m}\,\text{s}^{-1}$ \\
$V_X=v_x/U$ & Dimensionless horizontal velocity & -- \\
$V_s=v_s/U$ & Dimensionless interfacial velocity & -- \\
$V_Z=v_z/(\varepsilon U)$ & Dimensionless vertical velocity & -- \\
$A = \displaystyle{\frac{3\gamma_{\theta} \Delta T }{\rho g h_0^2}}$ & Dimensionless heating intensity & -- \\
$\text{Bi}=k h_0/\kappa$ &  Biot number & -- \\
$\text{Pe}_{\theta} = \ell_{\theta} U_{\theta} / \alpha$ & Thermal P\'eclet number & -- \\
$\text{Pe}_{s} = \ell_{\theta} U_{\theta} / D_s$ & Solutal P\'eclet number & -- \\
$\text{Ma}_s =\displaystyle{ \frac{\gamma_s \Gamma_0 h_0}{4\eta D_s}}$ & Marangoni number & -- 
\end{tabular}
\end{ruledtabular}
\caption{List of the physical parameters used in the text, with corresponding formula (if applicable) and SI units.}
\end{table}

We consider a thin film of a viscous liquid with viscosity $\eta$ and density $\rho$, which is bounded below by a solid surface. The latter defines the horizontal plane $z=0$, the $z$-axis being oriented upward. For the sake of simplicity, it is assumed that the system is invariant by translation along the $y$-coordinate. The system under study is thus effectively two-dimensional, as depicted in figure~\ref{fig1}. 

When the temperature is uniform, the liquid is still and the free surface remains flat. The situation changes when a localized heating is switched on: due to the temperature dependence of the surface tension~$\gamma$, a thermocapillary flow sets up and drives the liquid from the warmer regions to the colder regions. The resulting deformation  of the liquid-air interface is described by the height function $h(x,t)$ that we aim to characterize. 

The main difficulty when dealing with non-isothermal flows is that the physical properties of the liquid --- viscosity~$\eta$, mass density~$\rho$, surface tension~$\gamma$, heat capacity~$c$, heat conductivity~$\kappa$ --- are likely to depend on the local temperature~$T(x,z,t)$. In this paper, it is assumed that all material parameters are constant except for the surface tension in the Marangoni boundary condition. 

Momentum conservation in the liquid phase is enforced through the Navier-Stokes equation
\begin{equation}
\rho \left(    \partial_t \mathbf{v} + \mathbf{v} \cdot \bm{\nabla} \mathbf{v} \right) = \eta \nabla^2 \mathbf{v} - \bm{\nabla} p +\rho \mathbf{g}   \ ,  
\label{NS}
\end{equation} 
with $\mathbf{v}$ the velocity, $p$ the pressure, and $\mathbf{g}$ the gravitational acceleration. 
The Navier-Stokes equation has to be solved together with the incompressibility condition 
\begin{equation}
\bm{\nabla} \cdot \mathbf{v} = 0 \ .
\label{cont}
\end{equation} 
Regarding the boundary conditions, the no-slip and no-penetration requirements at the solid wall ($z=0$) are expressed as
$v_x(x,0,t)=0$ and $v_z(x,0,t)=0$, respectively.
At the free interface ($z=h$), the kinematic boundary condition reads
\begin{equation}
\mathbf{v}\cdot \mathbf{n}= \frac{\partial_t h}{\sqrt{1+(\partial_x h) ^2}} \ ,
\label{kinematic}
\end{equation}
while the stress continuity condition is 
\begin{equation}
\bm{\sigma} \cdot \mathbf{n} = \bm{\nabla}_{\parallel} \gamma -\gamma \left( \bm{\nabla} \cdot \mathbf{n} \right) \mathbf{n} \ .
\label{stress}
\end{equation}
Here, $\mathbf{n}$ is the unit normal vector pointing  toward the gas phase, and   $\bm{\nabla}_{\parallel}= \left( \bm{I} - \mathbf{n} \mathbf{n} \right) \cdot \bm{\nabla} $ is the surface gradient operator. The  components of the viscous stress tensor $\bm{\sigma}$ are written as $\sigma_{ij} = -(p-p_0) \delta_{ij} + \eta ( \partial_i v_j + \partial_j v_i)$, with $p_0$ the reference pressure in the gas phase. 
Finally, all fields are expected to decay to their equilibrium values far away from the disturbance 
\begin{subequations}
\label{bc_infinity}
\begin{align}
&\lim_{x\to \pm \infty} h(x,t)=h_0 \ ,  \label{bch}\\
&\lim_{x\to \pm \infty} T(x,z,t)=T_0 \ ,  \label{bct} \\
& \lim_{x\to \pm \infty} \mathbf{v}(x,z,t)=\bm{0} \ . \label{bcv}
\end{align}
\end{subequations}

We focus in this work on the limit of a thin liquid film.
More precisely, it is assumed that the film thickness $h_{0}$ is significantly smaller than the horizontal length scale~$\ell_{\theta}$ that characterizes the temperature gradient.  The set of equations~(\ref{NS})--(\ref{bc_infinity}) can then be solved within the standard lubrication approximation  $\varepsilon=h_0 /\ell_{\theta}\ll 1$. The derivation follows in every respect that given in the classical review article by Oron and colaborators~\cite{oronRMP1997}. At lowest order, the Navier-Stokes equation~(\ref{NS}) reduces to
\begin{equation}
\eta \partial_z^2 v_x = \partial_x p \ , \quad \text{and} \quad \partial_z p = -\rho g \ .
\label{NS_thinfilm}
\end{equation}
The stress continuity condition~(\ref{stress})  respectively gives rise to Laplace boundary condition $p(x,h,t)  = p_0 -\gamma \partial_x^2 h$ for the projection along the vertical  direction, and to Marangoni boundary condition $\eta \partial_z v_x \big\vert_{h} = \partial_x \gamma$ for the projection along the horizontal direction.
It is then straightforward to establish the expression of the horizontal velocity 
\begin{equation}
v_x(x,z,t) = \frac{\left(z^2-2zh\right)}{2\eta}  \partial_x \left(\rho g h- \gamma \partial_x^2 h \right) + \frac{z}{\eta} \partial_x \gamma  \ . 
\label{velocity_thinfilm}
\end{equation}
The derivation is completed by integrating the continuity equation~(\ref{cont}) from $z=0$ to $z=h$, together with the kinematic boundary condition~(\ref{kinematic}). One finally obtains the time evolution equation for the interface~\cite{oronRMP1997}
\begin{equation}
\partial_t h = \frac{1}{3\eta} \partial_x \left[ h^3 \partial_x \left( \rho g   h -\gamma  \partial^2_x  h \right) \right]   - \frac{1}{2 \eta}  \partial_x \left( h^2  \partial_x \gamma \right)  \ . 
\label{thinfilm}
\end{equation}
This general result is valid whatever the physical origin of the Marangoni flow, be it thermal or solutal.
Yet, it is necessary to specify the Marangoni stress $\partial_x \gamma$ in order to establish the time evolution equation for $h(x,t)$ in response to an external heating. This issue is addressed hereafter, first for a pure liquid, and subsequently for an interface loaded with surfactants.

\section{Temperature field and thermocapillary effect}
\label{sec_thermocap}

In the absence of surfactant, the surface tension  depends solely on the interfacial temperature, $\gamma=\gamma(T_s)$, with $T_s=T(x,h(x,t),t)$. For moderate deviation with respect to the equilibrium temperature $T_0$, we can assume a linear dependence of the surface tension
\begin{equation}
\gamma (T_s) = \gamma_0 - \gamma_{\theta} (T_s-T_0) \ , 
\label{gammaT}
\end{equation} 
where we define $\gamma_0=\gamma(T_0)$ and $\gamma_{\theta} = \left\vert \partial \gamma / \partial T \right\vert$.
Note that the surface tension is usually a decreasing function of the temperature. For most liquids, $\gamma_0$ lies in the range  $10^{-2}-10^{-1}~\text{N}~\text{m}^{-1}$, while $\gamma_{\theta}$ is typically of the order of~$10^{-4}~\text{N}~\text{m}^{-1}~\text{K}^{-1}$. The linear approximation~(\ref{gammaT}) is thus legitimate for a temperature increase up to 10~K.

Energy conservation in the liquid phase is then enforced by the heat equation
\begin{equation}
\rho c \left( \partial_t T + \mathbf{v} \cdot \bm{\nabla} T \right) = \kappa \nabla^2 T \ ,
\label{heat}
\end{equation} 
where the temperature decays to its equilibrium value at infinity according to~(\ref{bct}).
The way that heating is provided actually depends on the experimental system under consideration. Here, it is assumed that the temperature at the solid surface ($z=0$) is imposed 
\begin{equation}
T(x,0,t)= T_0 + \Delta T \vartheta(x) \ ,
\label{temp_surface}
\end{equation} 
with $\Delta T >0$, and $\vartheta(x)$ a function to be specified.  At the free interface ($z=h$),  Newton's law of cooling accounts for the  heat exchanges between the liquid phase and the gas phase
\begin{equation}
\kappa  \bm{\nabla} T \cdot \mathbf{n} + k \big( T_s-T_0 \big) =0  \ ,
\label{cooling}
\end{equation}
with $k$ the heat transfer coefficient. 

The heat equation~(\ref{heat}), together with the boundary conditions~(\ref{temp_surface}) and~(\ref{cooling}), are readily solved in the lubrication approximation~$\varepsilon \ll 1$. As shown in Appendix~A, equation~(\ref{heat}) reduces  to $\partial_z^2 T = 0$ at lowest order in $\varepsilon$. The temperature is therefore quasi-stationary and varies linearly with the distance to the substrate~\cite{oronRMP1997}. In the lubrication limit, Newton's law~(\ref{cooling}) also simplifies to
$\kappa \partial_z T \big\vert_{h} =- k \left( T_s-T_0 \right) \big\vert_{h} $.
The  temperature profile thus reads
\begin{equation}
T(x,z,t)=T_0 + \Delta T \vartheta(x) \left[\frac{h_0+\text{Bi}(h-z)}{h_0+\text{Bi}h}\right]  \ ,
\label{sol_temp}
\end{equation}
with $\text{Bi}=k h_0/\kappa$ the Biot number. Note that the time dependence is hidden in the film thickness~$h(x,t)$. The surface tension gradient immediately follows from the definition~(\ref{gammaT}) using the chain rule $\partial_x T_s=\partial_x T \vert_{h} + (\partial_x h) \partial_z T \vert_{h}$. The expression further simplifies in the limit $\text{Bi}\ll1$, and one eventually gets
\begin{equation}
\partial_x \gamma =  -\gamma_{\theta} \Delta T \partial_x \vartheta \ .
\label{grad_gammaT}
\end{equation}
Inserting this result in the general equation~(\ref{thinfilm}), one finally obtains the time evolution equation for the
deformation of a thin liquid film in a \textit{non-uniform} temperature field.
Note that this equation has already been derived previously, \textit{e.g.}, in the context of topography control of liquid interfaces~\cite{eshelFlow2022}. It provides a generalization of the classical evolution equation for a \textit{uniform} temperature~\cite{oronRMP1997}. From this equation, it is possible to explore for instance stability issues~\cite{koleskiPoF2020,wuPoF2024}, yet this point is not the focus of this article. In the following, we discuss the effect of insoluble surfactants on the shape of the interface.

\section{Competition between thermocapillary and solutocapillary flows}
\label{sec_solutocap}

We now assume that surface-active molecules are irreversibly adsorbed at the water-air interface, with surface concentration $\Gamma(x,t)$. In the absence of perturbation, the concentration is uniform: $\Gamma(x,t)= \Gamma_0$. When heating the system, the thermocapillary flow sets in and surfactants are advected away from the heat source.
The resulting concentration gradient then leads to a secondary flow, of solutal origin this time, which is directed in the opposite direction. This solutocapillary counterflow is expected to strongly reduce the deformation of the interface.

In general, the surface tension is a decreasing function of both the temperature and the concentration of surfactants. For small deviations with respect to equilibrium, one can assume a linear relationship
\begin{equation}
\gamma (T_s, \Gamma) = \gamma_0 - \gamma_{\theta} (T_s-T_0) - \gamma_s \left( \Gamma - \Gamma_0 \right) \ , 
\label{defgamma}
\end{equation} 
where the constant $\gamma_s$ is defined as $\gamma_s= \vert \partial \gamma / \partial \Gamma \vert$. In order to compute the surface tension gradient, one first needs to solve the advection-diffusion equation that governs the transport of surfactant molecules along a deformed interface~\cite{stonePoF1990}
\begin{equation}
\partial_t \Gamma  + \bm{\nabla}_{\parallel} \cdot \left(  \Gamma \, \mathbf{v}_{\parallel} \right)+\Gamma \left( \bm{\nabla}_{\parallel} \cdot \mathbf{n}\right) (\mathbf{v} \cdot \mathbf{n} )  = D_s \nabla_{\parallel}^2 \Gamma   \ ,
\label{advdiff}
\end{equation}
with $D_s$ the diffusion coefficient. Besides the standard diffusion and advection terms, equation~(\ref{advdiff}) also accounts for the variations of surfactant concentration as a result of the local changes in interfacial area.

In the lubrication approximation, equation~(\ref{advdiff}) simplifies right away since all the terms related to the local curvature of the interface are proportional to $\vert \partial_x h \vert^2$, and are therefore~$\sim \mathcal{O}(\varepsilon^2)$. Secondly, the surfactant layer is almost always surface incompressible, $\vert \Gamma - \Gamma_0 \vert \ll \Gamma_0$, even in the dilute regime~\cite{shardtJFM2016,manikantanJFM2020}. Assuming that the deviations from the average concentration are very small~\cite{bickelPRFluids2024}, it is shown in Appendix~B that the advection-diffusion equation~(\ref{advdiff}) eventually takes the very simple form
\begin{equation}
 \Gamma_0 \partial_x v_s   = D_s \partial_x^2 \Gamma \ ,
\label{advdiff_red}
\end{equation} 
with $v_s=v_x(x,h,t)$ the interfacial velocity. 
We can now proceed and integrate this equation with the appropriate limits $\Gamma(x,t) \to \Gamma_0$ and $\partial_x \Gamma \to 0$ when $x \to \pm \infty$. 
In particular, the concentration gradient is found to be directly proportional to the interfacial velocity: $\partial_x \Gamma = \Gamma_0 v_s /D_s$.
However, the interfacial velocity still needs to be made explicit. To this aim, we express the surface tension gradient  
from the combination of equations~(\ref{defgamma}) and~(\ref{advdiff_red}), and evaluate the horizontal velocity at the interface $z=h(x,t)$ using equation~(\ref{velocity_thinfilm}). This gives a closure equation for~$v_s$ --- see Appendix~C for the technical details.
Inserting the resulting expression of $\partial_x \Gamma$ in the original time evolution equation~(\ref{thinfilm})
ultimately leads to
\begin{widetext}
\begin{equation}
\partial_t h =   \frac{1}{3\eta}  \partial_x \left[   \left(\frac{1+ \text{Ma}_s \displaystyle{\frac{h}{h_0}}}{1+ 4\text{Ma}_s \displaystyle{\frac{h}{h_0}}} \right)  h^3 \partial_x \left( \rho g   h -\gamma_0  \partial^2_x  h \right) \right]  
 + \frac{\gamma_{\theta} \Delta T}{2\eta}  \partial_x \left( \frac{h^2  \partial_x \vartheta}{1+ 4\text{Ma}_s \displaystyle{\frac{h}{h_0}}} \right)  \ ,
\label{thinfilm_ts}
\end{equation}
\end{widetext}
where we define
\begin{equation}
\text{Ma}_s = \frac{\gamma_s \Gamma_0 h_0}{4\eta D_s}  \ .
\label{Ma}
\end{equation}
Equation~(\ref{thinfilm_ts}) is the central results of this article. It describes the hindered dynamics of a heated thin film in the presence of surfactants. The dimensionless parameter that controls the deformations is the solutal Marangoni number $\text{Ma}_s$, which is proportional to the surfactant concentration~$\Gamma_0$~\cite{manikantanJFM2020}. In the limit of vanishing concentration, $\text{Ma}_s\to 0$, one recovers the evolution equation for a surfactant-free interface. In the opposite limit $\text{Ma}_s \gg 1$, it is the second term on the right-hand side of equation~(\ref{thinfilm_ts}) --- the thermocapillary contribution ---  that undergoes the most significant modification due to the presence of surface-active molecules. Consequences regarding the  deformation profile are discussed in the following.

\section{Deformation profile in steady state}
\label{sec_deform}

In the stationary limit $\partial_t h = 0$, the thin film equation~(\ref{thinfilm_ts}) can  be integrated once (provided that a steady state exists). Since there is no flux at $x \to \pm \infty$, one obtains
\begin{equation}
\left(1+ \text{Ma}_s \displaystyle{\frac{h}{h_0}} \right)  h \partial_x \left( \rho g   h -\gamma_0  \partial^2_x  h \right)  =- \frac{3\gamma_{\theta} \Delta T}{2}    \partial_x \vartheta   \ .
 \label{thinfilm_stat}
\end{equation}
We can also express the interfacial velocity
\begin{equation}
v_s =  -  \left(1+ \text{Ma}_s \displaystyle{\frac{h}{h_0}} \right)^{-1}  h\frac{\gamma_{\theta} \Delta T}{4 \eta}  \partial_x \vartheta   \ .
 \label{velocity_stat}
\end{equation}
Note that $h(x)$ has the same parity as $\vartheta(x)$, while the parity of $v_s(x)$ is opposite.
Interestingly, one can directly  infer from~(\ref{velocity_stat}) that the interfacial velocity vanishes as $\text{Ma}_s^{-1}$ in the limit $\text{Ma}_s \gg 1$. The liquid interface becomes completely frozen when the concentration of surfactant is large.

To illustrate the discussion, we still need to specify the functional form of the reduced temperature~$\vartheta(x)$. Yet this choice is actually not critical: one only requires the function~$\vartheta(x)$ to be sufficiently smooth,  and such that its width~$\ell_{\theta}$ is much larger than the film thickness $h_0$. Three temperature profiles are considered in this work:
a Gaussian function
\begin{equation}
\vartheta(x)=e^{ -x^2/(2\ell^2_{\theta})}  \ ,
\label{gaussian}
\end{equation}
a Lorentzian function
\begin{equation}
\vartheta(x)=\frac{1}{1 + x^2/\ell_{\theta}^2} \ ,
\label{lorentz}
\end{equation}
and a positive, periodic function
\begin{equation}
\vartheta(x)=\cos^2 \left( \frac{x}{2 \ell_{\theta}}  \right) \ .
\label{periodic}
\end{equation}
All three functions are defined such that $\vartheta (0)=1$.

\begin{figure*}
\centering
\includegraphics[width=\textwidth]{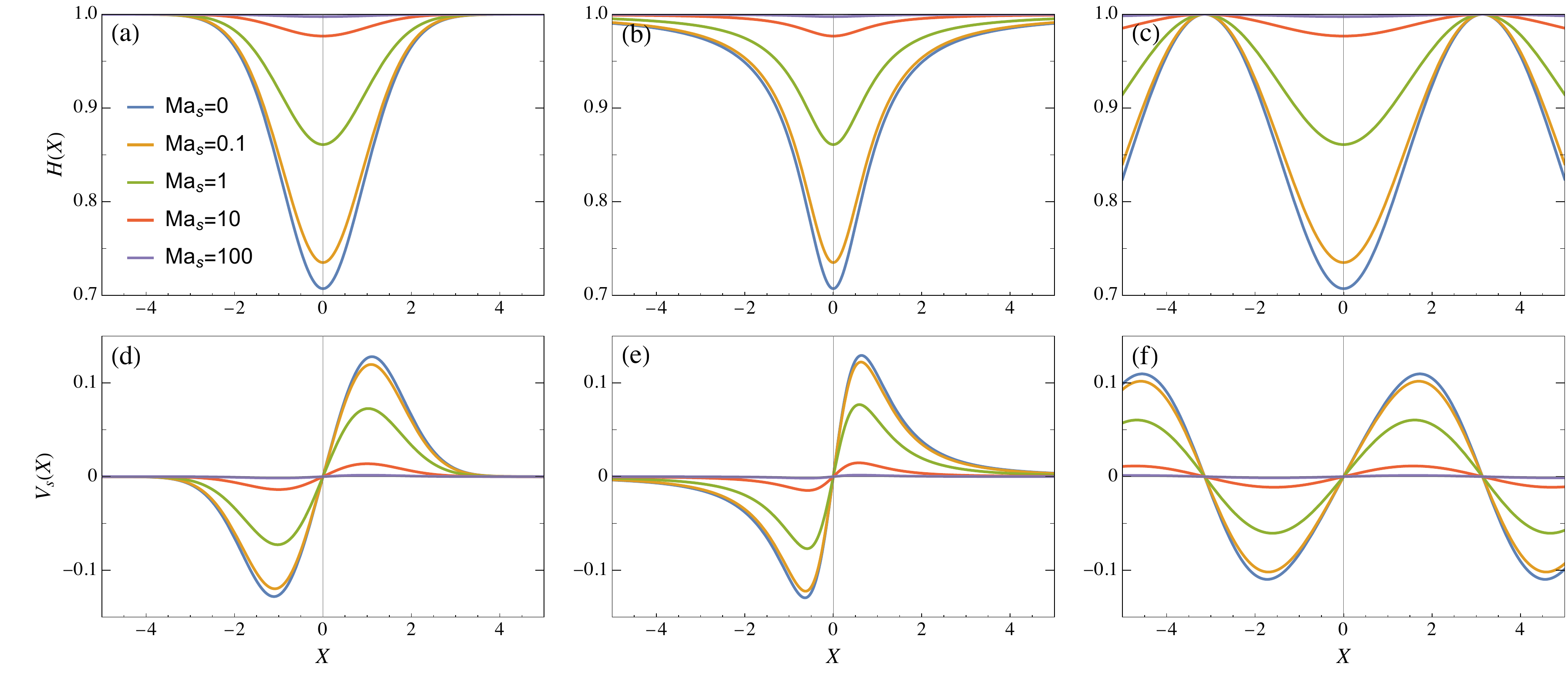}
\caption{Dimensionless deformation  (top) and dimensionless interfacial velocity (bottom) corresponding to a Gaussian, Lorentzian and periodic temperature  (from left to right), for different values of the Marangoni number. Other  parameters: $A=0.5$ and~$\lambda_c =0$.}
\label{fig2}       
\end{figure*}

\subsection{Dimensionless equations}

Equation~(\ref{thinfilm_stat}) being nonlinear, one has to resort either to approximate or to numerical methods. In both cases, it is appropriate to switch to dimensionless variables.
Given the geometry of the system, it is natural to define the dimensionless horizontal coordinate $X=x/\ell_{\theta}$ and interface height $H=h/h_0$, so that equation~(\ref{thinfilm_stat}) takes the dimensionless form
\begin{equation}
  \left(1+ \text{Ma}_s H \right)  H \partial_X \left( H -\lambda_c^2 \partial^2_X  H \right) 
 =- \frac{A}{2}    \partial_X \vartheta   \ .
 \label{thinfilm_stat_adim}
\end{equation}
Doing so, we introduced two new parameters. The first one, $A = 3\gamma_{\theta} \Delta T / (\rho g h_0^2)$, is the dimensionless parameter that controls the heating intensity and hence the strength of thermocapillary effect. The second parameter, $\lambda_c =\sqrt{\gamma_0/(\rho g \ell_{\theta}^2)}=\ell_c/ \ell_{\theta}$, is the dimensionless capillary length, expressed in units of~$\ell_{\theta}$. It characterizes the relative contribution of surface tension over gravitational forces. The concentration of surfactants, for its part, is specified by the Marangoni number $\text{Ma}_s$. 

Regarding the interfacial velocity~(\ref{velocity_stat}), it is scaled by the characteristic velocity of the problem $U$. The latter is determined by the balance between the viscous stress $\eta U/h_0$ and the Marangoni stress $\gamma_{\theta} \Delta T/l_{\theta}$. This leads to $U=\varepsilon U_{\theta}$, with $U_{\theta}=\gamma_{\theta} \Delta T/\eta$ the thermocapillary speed.
The dimensionless interfacial velocity $V_s=v_s/(\varepsilon U_{\theta})$ is then given by
\begin{equation}
V_s = -\frac{1}{4} \frac{H}{\left(1+\text{Ma}_s H\right)}    \partial_X \vartheta   \ .
 \label{vs_adim}
\end{equation}
Once the deformation profile $H$ is known, the interfacial velocity $V_s$ follows directly. The issue is then to  solve equation~(\ref{thinfilm_stat_adim}). Although the latter is nonlinear, it can be solved exactly in the gravitational regime $\lambda_c \ll 1$. We first focus on this limit before considering the more general case of finite $\lambda_c$.

\subsection{Gravitational regime: $\lambda_c \ll 1$}

Let us first consider the regime where surface tension can be neglected. In the limit $\lambda_c \ll 1$, equation~(\ref{thinfilm_stat_adim}) simplifies to
\begin{equation}
  \left(1+ \text{Ma}_s H \right)  H \partial_X H
 =- \frac{A}{2}    \partial_X \vartheta   \ .
 \label{thinfilm_stat_adim_grav}
\end{equation}
Taking into account the boundary condition $H(X) \to 1$ when $X \to \pm \infty$, the deformation profile is obtained as
\begin{equation}
\frac{2\text{Ma}_s}{3} H(X)^3 + H(X)^2
 =1 +  \frac{2\text{Ma}_s}{3}- A \vartheta (X)   \ .
 \label{sol_grav_adim}
\end{equation}
Even though this implicit expression could be made explicit, it leads to a complicated expression that is not very useful. Instead, let us discuss the solutions corresponding to the two asymptotic regimes $\text{Ma}_s \ll 1$ and $\text{Ma}_s \gg 1$. 

In the limit of vanishing Marangoni number, $\text{Ma}_s \ll 1$, the interface is devoid of surfactant and the deformation profile is given by
\begin{equation}
H(X) =\sqrt{1 - A \vartheta (X)}   \ .
\label{sol_grav_adim_0}
\end{equation}
Since we defined the reduced temperature such that $\vartheta(0)=1$, one can notice that the stationary solution only exists for $A \leq 1$. If the heating is too strong, gravitational forces cannot balance the thermocapillary stresses.  The film is therefore unstable when $A>1$, unless short range repulsive forces (\textit{e.g.}, van der Waals, not accounted for here) possibly become relevant at very small length scales~\cite{clavaudPRL2021}.

In the opposite limit of large surfactant concentration, $\text{Ma}_s \gg 1$, a brief inspection of equation~(\ref{sol_grav_adim}) reveals that the deformations of the interface are suppressed in the limit $\text{Ma}_s \to \infty$. To characterize the asymptotic behavior, we can look for a series expansion of the form $H(X)=1 + \text{Ma}_s^{-1}H^{(1)}(X)+ \mathcal{O}\left( \text{Ma}^{-2}_s \right)$. One readily obtains at lowest order
\begin{equation}
H(X) =1 - \frac{A}{2\text{Ma}_s} \vartheta (X)    \ ,
\label{sol_grav_adim_inf}
\end{equation}
showing that the deformation amplitude vanishes as~$H\sim \text{Ma}_s^{-1}$ when the concentration of surfactants becomes large.

For intermediate values of the Marangoni number, equation~(\ref{sol_grav_adim}) is solved numerically. We show on the top row of figure~\ref{fig2}   the evolution of the shape of the interface when the solutal Marangoni number is varied, for a fixed value of the heating power (here arbitrarily set to $A=0.5$). The three rows correspond to the three temperature profiles defined by equations~(\ref{gaussian}), (\ref{lorentz}) and~(\ref{periodic}), respectively. As expected from the asymptotic analysis, the amplitude of the deformation is strongly reduced as the Marangoni number increases.  
One can also notice that the three plots present the same general features and depends only moderately on the actual functional form of the temperature profile.

The accompanying flow in the liquid layer is also strongly suppressed as the surfactant concentration increases. The interfacial velocity $V_s$, which is obtained from equation~(\ref{vs_adim}), is shown on the bottom row of figure~\ref{fig2}. In the absence of surfactant ($\text{Ma}_s = 0$), the velocity presents an extremum at  $X_m \approx \pm 1$. The position of this maximum hardly varies in the presence of surfactant, yet its amplitude decays with the same scaling $V_s \sim \text{Ma}_s^{-1}$ for large values of the Marangoni number. We also observe a similar velocity profiles for the three cases considered in this work.

\begin{figure*}
\centering
\includegraphics[width=\textwidth]{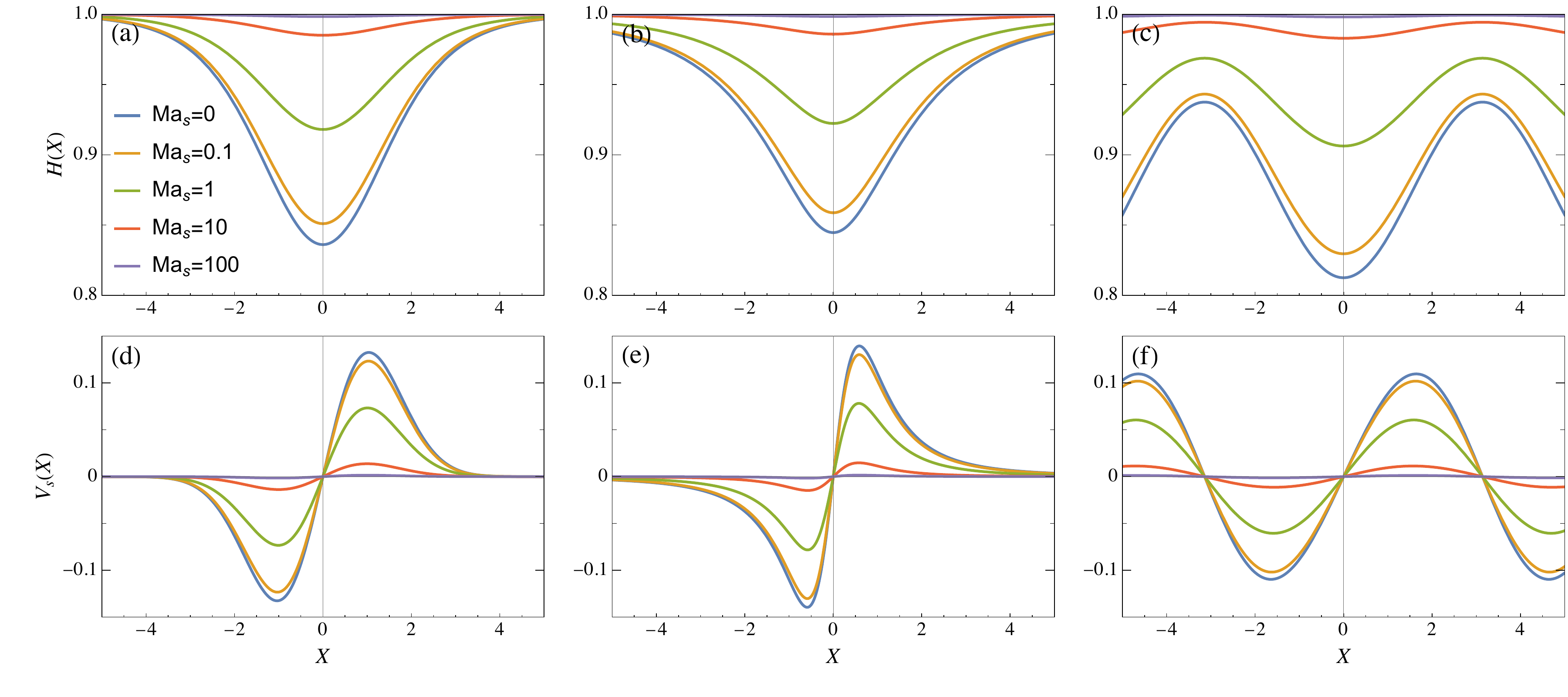}
\caption{Dimensionless deformation  (top) and dimensionless interfacial velocity (bottom) corresponding to a Gaussian, Lorentzian and periodic temperature  (from left to right), for different values of the Marangoni number. Other  parameters: $A=0.5$ and~$\lambda_c =1$.}
\label{fig3}       
\end{figure*}

\subsection{Effect of capillarity}

If the reduced capillary length $\lambda_c$ is finite, the deformation is generally further reduced compared to the gravitational regime $\lambda_c=0$, even in the absence of surfactant. It is thus relevant to seek an approximate solution of the problem. We then write $H(X)=1+\delta H(X)$, with $\delta H \ll H$, and develop the differential  equation~(\ref{thinfilm_stat_adim}) up to linear order. The resulting equation is readily integrated once to give
\begin{equation}
   \lambda_c^2 \partial^2_X  \delta H - \delta H
 = \frac{A}{2 \left(1+ \text{Ma}_s \right)}   \vartheta (X)   \ .
 \label{thinfilm_stat_adim_delta}
\end{equation}
This equation is linear and can be solved for any temperature profile using Green function techniques. The solution then reads
\begin{equation}
\delta H(X) = \frac{A}{2 \left(1+ \text{Ma}_s \right)} \int_{-\infty}^{+\infty}  \vartheta (X') \mathcal{G} (X-X') \d X'  \ ,
\label{convol}
\end{equation}
where the Green function is given by
\begin{equation}
\mathcal{G}(X) = -\frac{1}{2 \lambda_c} e^{-\vert X \vert / \lambda_c } \ .
\label{green}
\end{equation}
The convolution integral in equation~(\ref{convol}) can then be calculated together with the Green function~(\ref{green}) and the functional form of the temperature, either  given by~(\ref{gaussian}), (\ref{lorentz}) or~(\ref{periodic}).

Figure~\ref{fig3} shows the deformation profiles and interfacial velocities for $\lambda_c=1$, all things being equal. Compared to figure~\ref{fig2}, one can notice that the amplitude of the deformation is reduced (remark that the vertical scale is different). At the same time, the width of the deformation is increased. The case of the periodic temperature [figure~\ref{fig3}(c)] appears somewhat peculiar: whereas the deformation would vanish at each value of $X_k=(2k+1)\pi$ (with $k$ an integer) in the gravitational regime, this is not the case anymore when surface tension is present. Indeed, the deformation profile obtained by integrating~(\ref{convol}) for the periodic temperature~(\ref{periodic}) is 
\begin{equation}
H(X)=1- \frac{A}{4\left(1+ \text{Ma}_s \right)}\left[ 1 + \frac{\cos X}{1 + \lambda_c^2} \right] \ .
\end{equation}
Consequently, one has $H(X_k)\neq 1$ as soon as $\lambda_c >0$, and the difference actually increases with $\lambda_c$. As expected, surface tension tends to smooth out the deformations of the interface.

To better understand the competition between gravitational and capillary effects, we focus henceforth on the Gaussian temperature  given by equation~(\ref{gaussian}). In this case, the integral~(\ref{convol}) can be evaluated exactly and one obtains the deformation profile 
\begin{widetext}
\begin{equation}
H(X) = 1 -  \frac{A}{4  \lambda_c \left(1+ \text{Ma}_s \right)}  \sqrt{\frac{\pi}{2}}e^{1/(2\lambda_c^2)} \left[  \exp\left( \frac{X}{\lambda_c}\right) \text{erfc} \left( \frac{1}{\lambda_c \sqrt{2}} +  \frac{X}{\sqrt{2}}  \right)  + \exp\left(- \frac{X}{\lambda_c}\right) \text{erfc} \left( \frac{1}{\lambda_c \sqrt{2}} -  \frac{X}{\sqrt{2}}  \right)  \right]  \ .
 \label{thinfilm_approx_gaussian}
\end{equation}
\end{widetext}

\begin{figure}
\centering
\includegraphics[width=\columnwidth]{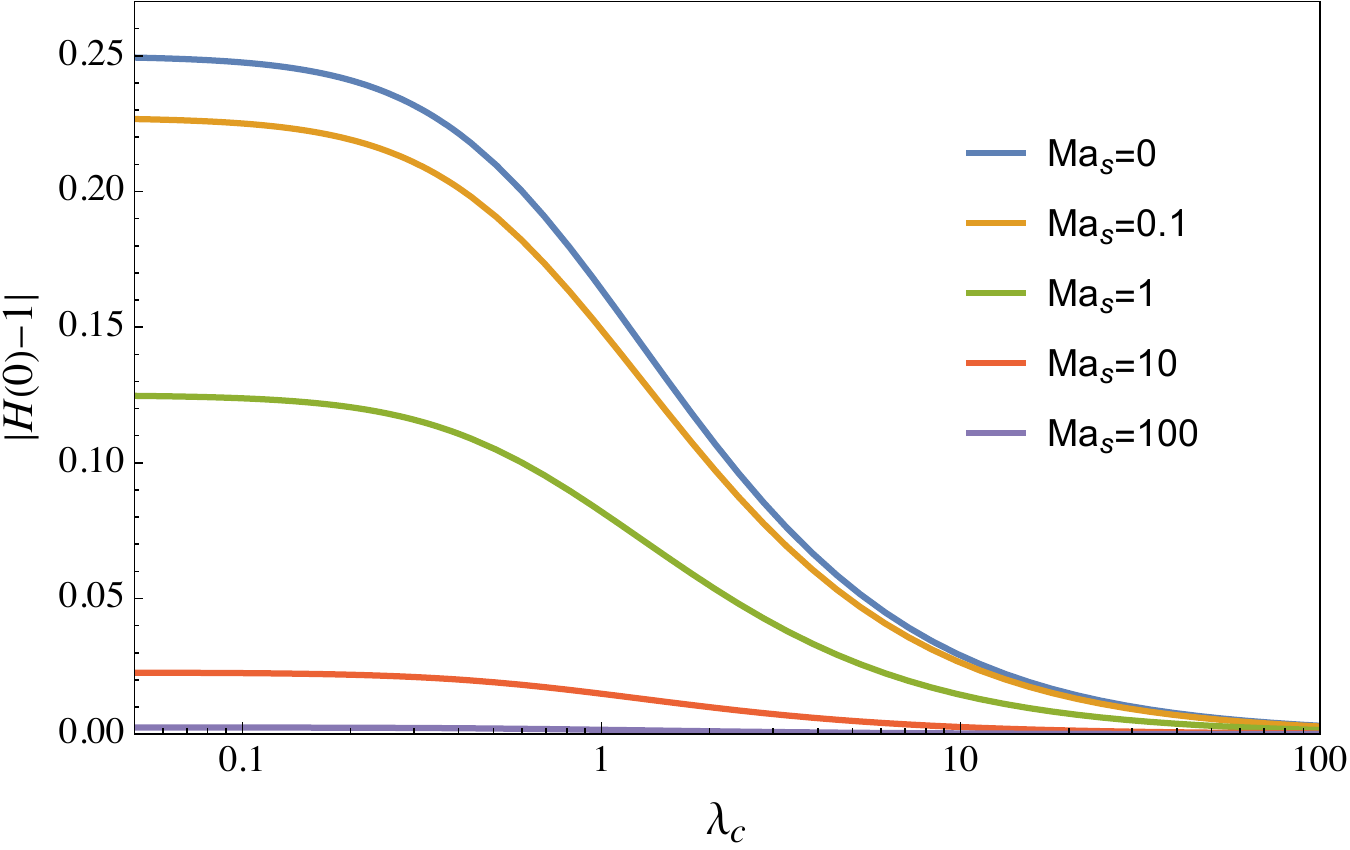}
\caption{Maximum dimensionless deformation corresponding to the Gaussian temperature as a function of the reduced capillary length, for different values of the Marangoni number. Other  parameter: $A=0.5$.}
\label{fig4}       
\end{figure}

\begin{figure}
\centering
\includegraphics[width=\columnwidth]{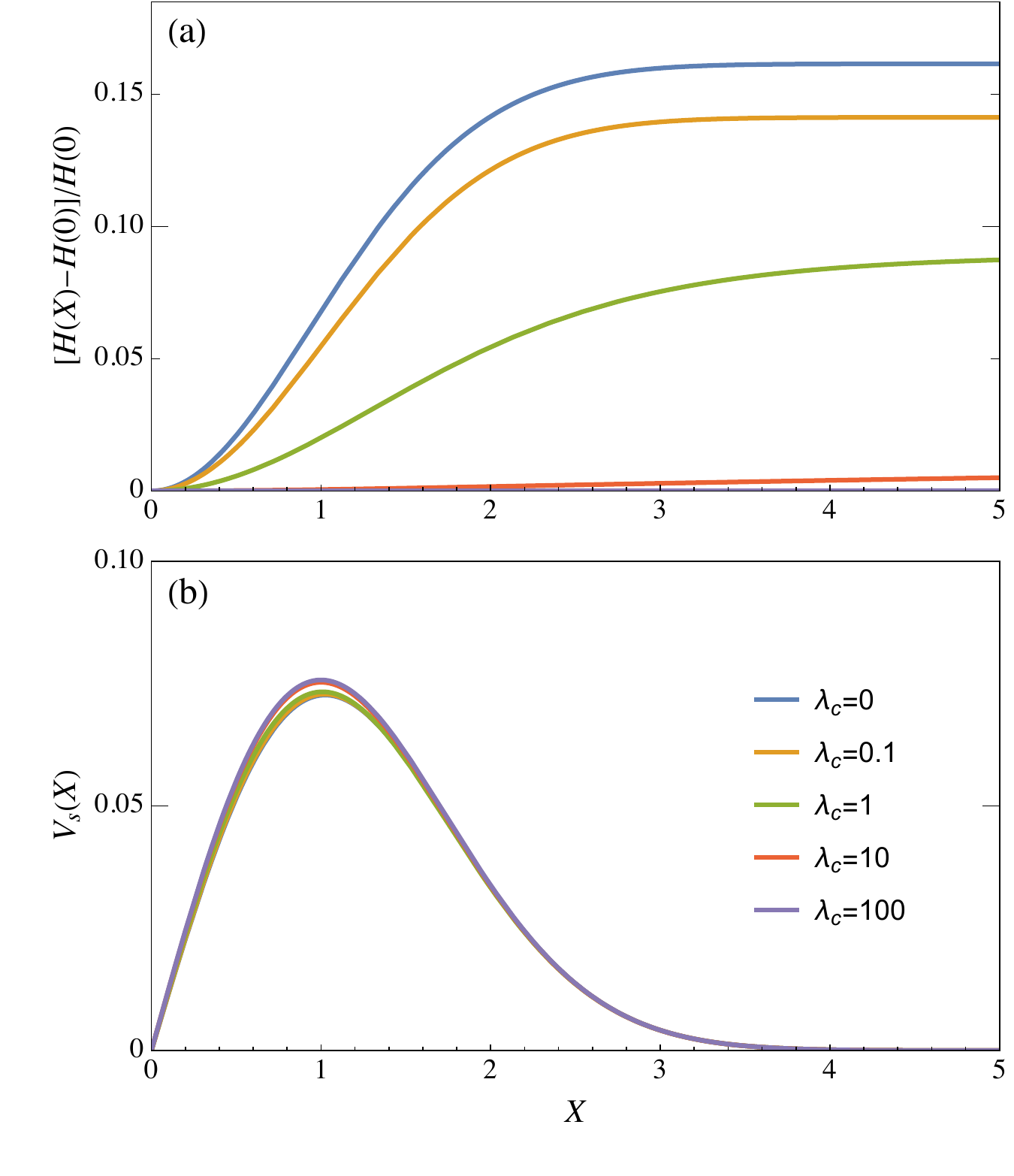}
\caption{Relative dimensionless deformation (top) and dimensionless interfacial velocity (bottom) corresponding to a Gaussian temperature, for different values of the reduced capillary length. Other  parameters: $A=0.5$ and~$\text{Ma}_s =1$.}
\label{fig5}       
\end{figure}

We first plot  in figure~\ref{fig4} the maximum deformation $\vert H(0)-1 \vert$  as a function of $\lambda_c$, in log scale. For any given value of  the Marangoni number, $\vert H(0)-1 \vert$ is a decreasing function of~$\lambda_c$. This is expected as increasing the capillary length amounts to increasing the surface tension. To proceed further, we show in figure~\ref{fig5}(a) the relative deformation $[H(X)-H(0)]/H(0)$.
Here, the value of the Marangoni number is arbitrarily set to $\text{Ma}_s = 1$. One can notice that the width of the deformation profile increases significantly when increasing~$\lambda_c$. As a matter of fact, the asymptotic expression of~(\ref{thinfilm_approx_gaussian}) in the limit $\lambda_c \gg 1$ reads
\begin{align}
H(X) & = 1 - \frac{A }{4 \lambda_c \left(1+ \text{Ma}_s \right)} \sqrt{2 \pi}e^{1/(2\lambda_c^2)} \nonumber \\
&\times \left[  \cosh \left( \frac{X}{\lambda_c}\right)  -\text{erf} \left(  \frac{X}{\sqrt{2}}  \right)  \sinh \left(\frac{X}{\lambda_c}   \right)  \right]  \ .
\end{align}
This expression reveals that the width of the profile is  primarily set by the capillary length, rather than the thermal length, in the tension-dominated regime.

More surprisingly, figure~\ref{fig5}(b) indicates that the interfacial velocity is almost unaffected by surface tension. This point can already be noticed if one compares  figures~\ref{fig2}(d)--(f) and~\ref{fig3}(d)--(f). To explain these observations, one has to focus on the general expression~(\ref{vs_adim}) of the interfacial velocity that reads $V_s(X) \approx -\frac{1}{4} H(X) \partial_X \vartheta$ to a first approximation. For the Gaussian temperature profile, $\partial_X \vartheta$ vanishes at $X=0$. The fact that the deformation amplitude is maximum at $X=0$ is therefore irrelevant. On the other hand,  $\partial_X \vartheta$ exhibits two extrema at $X_m=\pm 1$.  But even though the amplitude of the deformation decreases with~$\lambda_c$, its width  on the contrary increases with~$\lambda_c$. These two antagonist effects almost cancel each other in such a way that, for a given position~$X$ around~$X_m$, $H(X)$ is practically independent of~$\lambda_c$. It only increases by a few percents over several decades of~$\lambda_c$, which is barely visible in figure~\ref{fig5}(b).

\section{Discussion}
\label{sec_disc}

To summarize, we have studied the competition between thermocapillary and solutocapillary flows in a thin liquid film. The overall deformation of the interface is driven by the thermocapillary effect, while the presence of insoluble surfactants counteracts
the primary flow. Both the resulting flow and the  deformation can be suppressed if the Marangoni number $\text{Ma}_s= \gamma_s \Gamma_0 h_0/(4\eta D_s)$, which is proportional to  the concentration of surfactant, is sufficiently large.

We can  deduce from our analysis the concentration from which this effect becomes significant.
Taking the typical values $D_s \approx 10^{-9}~\text{m}^2~\text{s}^{-1}$, $\eta \approx 10^{-3}~\text{Pa}~\text{s}$, $\gamma_s\approx 10^{21}~\text{J}$~\cite{manikantanJFM2020}, and a film thickness $h_0 \approx 10^{-4}~\text{m}$, one finds that $\text{Ma}_s \approx 100$ for a surface concentration as low as $\Gamma_0 \approx 10^3$~molecules per $\mu \text{m}^{2}$. Interestingly, this estimate coincides with previous experimental or numerical observations that were obtained in different contexts~\cite{huJPCB2006,bickelPRF2019}.

We emphasize that our general method can be adapted in a straightforward manner to the various experimental configurations that can be envisaged. For instance, heating can be provided in the bulk (using a laser beam or a thin electric layer) rather than at the solid surface. In this case,  a source term need to be included in the heat equation~(\ref{heat}), which might change the functional form of~$\vartheta(x)$. But no major modification is expected otherwise. Also, extension to the 3D axisymmetric geometry should not present any significant difficulty. Finally, it could also be relevant to increase the complexity of the model by considering for instance the situation where the Biot number is finite. In this case, one expects an intricate coupling between the interface shape and the temperature field, that would deserve a specific study.

To conclude, we have developed a versatile approach that allows to characterize the reduction of the thermocapillary effect due to  the elastic response of the surfactant layer. While the presence of surfactants is generally difficult to assess, our predictions must provide a concrete and effective method for collecting quantitative information regarding surfactant concentration.

\section*{acknowledgments}

The authors wish to thank the International Research Network  ``Hydrodynamics at small scales'', under the auspice of CNRS, for the organization of the workshop held in IIT Madras, India on 25-27 October 2024, and where this work was first presented.

\section*{Author declarations}

\subsection*{Conflicts of interest}

The authors have no conflicts to disclose.

\subsection*{Author Contributions}

\noindent \textbf{D.~Kumar}: Formal analysis (lead); Investigation (equal); Writing – review \& editing (equal). \textbf{P.K.~Panigrahi}: Funding acquisition (lead); Project administration (equal); Writing – review \& editing (equal). \textbf{T.~Bickel}: Conceptualization (lead); Investigation (equal); Project administration (equal); Writing – original draft (lead); Writing – review \& editing (equal).

\subsection*{Data availability}

The data that support the findings of this study are available from the corresponding author upon reasonable request.

\appendix

\section{Solution of heat equation}

In this appendix, we establish the expression of the temperature field given in equation~(\ref{sol_temp}). The derivation closely follows that of reference~\cite{oronRMP1997}. We start by introducing the dimensionless variables $X=x/l_{\theta}$ and $Z=z/h_0$, with $\varepsilon=h_0/\ell_{\theta}$.  The components of the velocity read $V_X=v_x/U$ and $V_Z=v_z/(\varepsilon U)$, as required by the continuity condition. The velocity scale $U$ is set by  $U =\varepsilon U_{\theta}$, with $U_{\theta}=\gamma_{\theta}\Delta T/\eta$ the thermocapillary speed. The dimensionless time $\tau$ is expressed in units of the advection time scale $\ell_{\theta}/U$. The reduced temperature is defined as $\Theta = (T-T_0)/\Delta T$. Then rewriting the heat equation~(\ref{heat})   in dimensionless form, one gets
\begin{equation}
\varepsilon^3 \text{Pe}_{\theta} \left( \partial_{\tau} \Theta + V_X \partial_X \Theta +V_Z \partial_Z \Theta \right) = \varepsilon^2 \partial_X^2 \Theta + \partial_Z^2 \Theta \ ,
\label{heat_adim}
\end{equation} 
with $\text{Pe}_{\theta}=\ell_{\theta} U_{\theta} / \alpha$ the thermal P\'eclet number, and $\alpha= \kappa/(\rho c)$ the heat diffusion coefficient.
In the lubrication approximation $\varepsilon \ll 1$, the heat equation simplifies to  
\begin{equation}
\partial_Z^2 \Theta=0 \ .
\end{equation} 
The temperature is therefore quasi-stationary and varies linearly with the distance to the substrate
\begin{equation}
\Theta(X,Z,\tau)=A(X,\tau)Z + B(X,\tau) \ .
\end{equation} 
To determine $A$ and $B$, we express the boundary conditions~(\ref{temp_surface}) and~(\ref{cooling}), that respectively read in dimensionless form $\Theta(X,0,\tau)=\vartheta(X)$ and $\partial_Z \Theta \vert_H=-\text{Bi} \varTheta \vert_H$.
The solution is then 
\begin{equation}
\Theta(X,Z,\tau)=\vartheta(X) \left[\frac{1+\text{Bi}(H-Z)}{1+\text{Bi}H}\right]  \ ,
\end{equation}
where the time dependence is hidden in the film thickness~$H(X,\tau)$.
Restoring the dimensions, we obtain the temperature profile
\begin{equation}
T(x,z,t)=T_0 + \Delta T \vartheta(x) \left[\frac{h_0+\text{Bi}(h-z)}{h_0+\text{Bi}h}\right]  \ ,
\end{equation}
In particular, the interfacial temperature $T_s=T(x,h,t)$ is given by
\begin{align}
T_s(x,t) & =T_0 + \Delta T \vartheta(x) \left(1+\text{Bi}\frac{h}{h_0}\right)^{-1}  \nonumber \\
& = T_0 + \Delta T \vartheta(x) + \mathcal{O}\left( \text{Bi} \right) \ .
\end{align}
In the limit of vanishingly small Biot number $\text{Bi} \ll 1$, the interfacial temperature is thus stationary and is entirely set by the temperature at the solid substrate.

\section{Advection-diffusion equation}

We proceed likewise for the advection-diffusion equation~(\ref{advdiff}). First, since $\vert \partial_x h \vert \sim \mathcal{O}(\varepsilon) \ll 1$, all terms that are related to the local curvature are of order $\varepsilon^2$ and can therefore be neglected. Equation~(\ref{advdiff}) then simplifies to 
\begin{equation}
\partial_t \Gamma  + \partial_x \left(  \Gamma \, v_s \right) = D_s \partial_x^2 \Gamma   \ .
\label{adv_diff_app}
\end{equation}
To proceed further, this equation is rewritten in dimensionless form.
Assuming that the surfactant layer is almost incompressible, we define the dimensionless concentration $C = (\Gamma-\Gamma_0)/(\varepsilon \Gamma_0)$. The other dimensionless variables are defined according to Appendix~A. If these dimensionless variables are substituted into~(\ref{adv_diff_app}), one obtains 
\begin{equation}
\text{Pe}_s \left[ \varepsilon \partial_{\tau} C  + \partial_X V_X + \varepsilon \partial_X \left(  CV_X \right)\right] =  \partial_X^2 C   \ ,
\end{equation}
where we define a second P\'eclet number, of solutal origin this time, $\text{Pe}_s = \ell_{\theta} U_{\theta}/D_s$. Keeping only the zeroth-order terms in the limit $\varepsilon \ll 1$, the advection-diffusion equation simplifies to
\begin{equation}
\text{Pe}_s  \partial_X V_X =  \partial_X^2 C   \ .
\end{equation}
Coming back to dimensional variables, we finally obtain
\begin{equation}
 \Gamma_0 \partial_x v_s   = D_s \partial_x^2 \Gamma \ ,
\end{equation} 
which corresponds to equation~(\ref{advdiff_red}) in the text.

\section{Derivation of equation~(\ref{thinfilm_ts})}

Equation~(\ref{thinfilm_ts}) being the central result of this article, let us give some details regarding its derivation. We start by expressing the interfacial velocity $v_s(x,t)=v_x(x,h,t)$ which follows from equation~(\ref{velocity_thinfilm})
\begin{equation}
v_s(x,t) = -\frac{h^2}{2\eta}  \partial_x \left(\rho g h- \gamma \partial_x^2 h \right) + \frac{h}{\eta} \partial_x \gamma  \ . 
\label{c1}
\end{equation}
The surface tension gradient is obtained by differentiating~(\ref{defgamma}), so that
\begin{equation}
\partial_x \gamma =  - \gamma_{\theta} \Delta T \partial_x \vartheta - \gamma_s \partial_x \Gamma  \ . 
\label{c2}
\end{equation} 
The last term is related to the interfacial velocity through the advection-diffusion equation~(\ref{advdiff_red}), which leads to
\begin{equation}
\partial_x \Gamma = \Gamma_0 \frac{v_s }{D_s}  \ . 
\label{c3}
\end{equation} 
Bringing~(\ref{c1}), (\ref{c2}) and~(\ref{c3}) together, we get a closure relation for the interfacial velocity
\begin{equation}
v_s  =  -  
\frac{ \frac{h^2}{2\eta} \partial_x \left(\rho g h- \gamma \partial_x^2 h \right) -\frac{h}{\eta} \gamma_{\theta} \Delta T  \partial_x \vartheta }{1+ 4\text{Ma}_s  \frac{h}{h_0}}   \ ,
\end{equation}
where the expression of the Marangoni number $\text{Ma}_s$ is defined in~(\ref{Ma}).
The surface tension gradient then follows from equations~(\ref{c3}) and~(\ref{c2}). Finally, inserting in the general time evolution equation~(\ref{thinfilm}), one obtains after some straightforward algebra the desired result.

\end{document}